\documentclass[sigconf,screen]{acmart}
\usepackage{graphicx}
\usepackage{xcolor}
\usepackage{pifont}
\usepackage{multirow}
\usepackage{makecell}
\usepackage{tcolorbox}
\usepackage{enumitem}
\usepackage{booktabs}
\usepackage{todonotes}
\usepackage{subcaption}
\usepackage[utf8]{inputenc}
\usepackage{textcomp}

\newboolean{showcomments}
\setboolean{showcomments}{true} 
\ifthenelse{\boolean{showcomments}}
 { \newcommand{\mynote}[2]{
      \fbox{\bfseries\sffamily\scriptsize#1}
        {\small$\blacktriangleright$\textsf{\emph{#2}}$\blacktriangleleft$}}}
        { \newcommand{\mynote}[2]{}}

\renewcommand{\shortauthors}{Alberick Euraste Djire}



\setcopyright{cc}
\setcctype{by}
\acmDOI{10.1145/3832783.3844566}
\acmYear{2026}
\copyrightyear{2026}
\acmISBN{979-8-4007-2882-2/2026/10}
\acmConference[ASE '26]{Proceedings of the 41st IEEE/ACM International Conference on Automated Software Engineering}{October 12--16, 2026}{Munich, Germany}
\acmBooktitle{Proceedings of the 41st IEEE/ACM International Conference on Automated Software Engineering (ASE '26), October 12--16, 2026, Munich, Germany}
\acmSubmissionID{ase26src-p23-p}
\received{2026-07-31}
\received[accepted]{2026-08-25}

\begin{document}

\title{Code Detectors Have a Half-Life: Obsolescence and Metric Illusions in LLM-Generated Code Detection}

\author{Alberick Euraste Djire}
\correspondingauthor
\orcid{0009-0007-0406-9573}
\affiliation{%
  \institution{University of Luxembourg}
  \city{Luxembourg}
  \country{Luxembourg}
}
\email{euraste.djire@uni.lu}

\begin{abstract}
Code detectors can become obsolete as code-generating models
evolve: a detector validated on one generation of models may not
transfer to the next. We call this limited useful life a detector half-
life. We evaluate eight general-purpose LLM judges and three dedicated detectors on human-written code and code produced by seven generators across C++, Java, and Python. Our results reveal two problems. First, performance varies considerably across generators and prompting strategies, suggesting that some detectors rely on generator-specific patterns rather than general evidence of code provenance. Second, accuracy can conceal severe prediction bias. DetectCodeGPT and GPT-Sniffer achieved an accuracy of 0.50 but an $F_1$ score of 0.00 across all generators because they classified almost every sample as AI-generated. However, general-purpose LLM judges achieved stronger accuracy and $F_1$ scores. Our results show that general-purpose LLMs are promising training-free judges of code provenance and can outperform dedicated detectors. However, their reliability depends on the judge model, the code generator, and the prompting strategy. We therefore recommend evaluating LLM judges across multiple generators and reporting macro-$F_1$ alongside class-specific precision and recall.
\end{abstract}

\maketitle

\section{Introduction}
\label{sec:introduction}

LLMs now generate, complete, and refactor source code~\cite{zan_large_2023, hou_large_2024}. Undisclosed use can undermine educational assessment, complicate license and intellectual-property review, and make security-sensitive code harder to vet, as LLMs can reproduce insecure patterns from their
training data. Reliable code-provenance detection is therefore increasingly important. However, code is harder to classify than natural language because syntactic constraints compress the token-distribution differences used by text detectors~\cite{shi_between_2024, ye_uncovering_2024}. Existing detectors commonly require white-box access, labeled data, or dedicated infrastructure~\cite{yang_zero-shot_2023, nguyen_gptsniffer_2024, jianwang_empirical_2024}. More fundamentally, they are validated against a bounded set of generators. As generator families and output styles evolve, reported performance may expire a limitation we call \emph{detector half-life}. Furthermore, the common reliance of detectors on accuracy creates a second risk: on a balanced dataset, a detector that predicts only one class still scores 0.50.

General-purpose LLMs offer a simple alternative. Instead of training a separate
classifier, an LLM can be asked whether a program was written by a human or
generated by another model. This makes LLM judges attractive as training-free
detectors. However, LLM judgments can be affected by model preferences,
prompting choices, and superficial code features \cite{pan_assessing_2024,
moon2026dont}.

This paper studies both the promise and the limits of LLM judges. We compare
eight general-purpose LLMs with three dedicated detectors across seven code
generators and three programming languages. Specifically, we ask how accurately general-purpose LLMs distinguish AI-generated from human-written code without task-specific training (\textbf{RQ1}), and how lexical, structural, and lexico-semantic guidance affect their accuracy and decision biases (\textbf{RQ2}). We make the following contribution. \ding{182} evidence that current code detectors have a half-life; \ding{183} an empirical demonstration that accuracy can mask detector collapse; and \ding{184} a cross-generator evaluation of LLM judges as a more adaptable, though still imperfect, alternative.

\section{Related Works}
\label{sec:related_work}

LLM-output detection uses trained classifiers, zero-shot statistics, or watermarking~\cite{nguyen_gptsniffer_2024, yang_zero-shot_2023, li_resilient_2024, mitchell2023detectgpt}. Code-specific variants add stylistic, task-conditioned, finetuned, or signal-based features~\cite{shi_between_2024, ye_uncovering_2024, skurla2026mcdok, yao2026coderipple}. These methods are typically tested on a fixed set of existing generators and reported mainly using accuracy, leaving their robustness to newer generators and susceptibility to class collapse unclear.

General-purpose LLM evaluators avoid detector training~\cite{pan_assessing_2024} and may adapt as foundation models improve, though prompt and judge biases remain~\cite{moon2026dont}. 
Unlike ~\citet{pan_assessing_2024}, who evaluate a single prompting strategy, we systematically compare direct and feature-guided prompting across eight judges, seven generators, and three languages, revealing that prompting strategy interacts with judge identity in ways a single-condition evaluation cannot expose.

\section{Methodology}
\label{sec:methodology}

We use the AI-Human-Generated-Program-Code Dataset~\cite{idrisov_program_2024}: 18 programming problems (C++, Python, Java), each paired with one human-written solution and one AI-generated solution per generator (StarCoder, ChatGPT, Bing AI Chat, Code Llama, CodeWhisperer, GitHub Copilot, InstructCodeT5+), yielding 18 human and $18{\times}7{=}126$ AI snippets, with the same 18 human solutions used as the baseline for every generator.
We evaluate eight LLM judges (Llama~3.1-70B, Gemini-2.5-Flash, Gemini-3-Flash, Ministral-14B, Mistral Large, GPT-3.5-Turbo, GPT-5-mini, Claude Sonnet~5) at temperature $0.1$, returning a JSON verdict (\texttt{"yes"} = LLM-generated, \texttt{"no"} = human-written) with justification; human-written code is the positive class. Each query is repeated three times and averaged.
We assess four prompting strategies: \ding{182} \emph{Direct} zero-shot classification; \ding{183} \emph{Lexical-guided}, extracting comments, identifiers, keywords, and libraries first; \ding{184} \emph{Structural-guided}, examining control-flow and formatting first; and \ding{185} \emph{Lexico-semantic-guided}, combining lexical cues with inferred purpose and semantics first.

\section{Experiments and Results}
\label{sec:experiments}

\begin{figure}[t]
    \centering
    \begin{subfigure}[b]{0.48\textwidth}
         \centering
         \includegraphics[width=\textwidth]{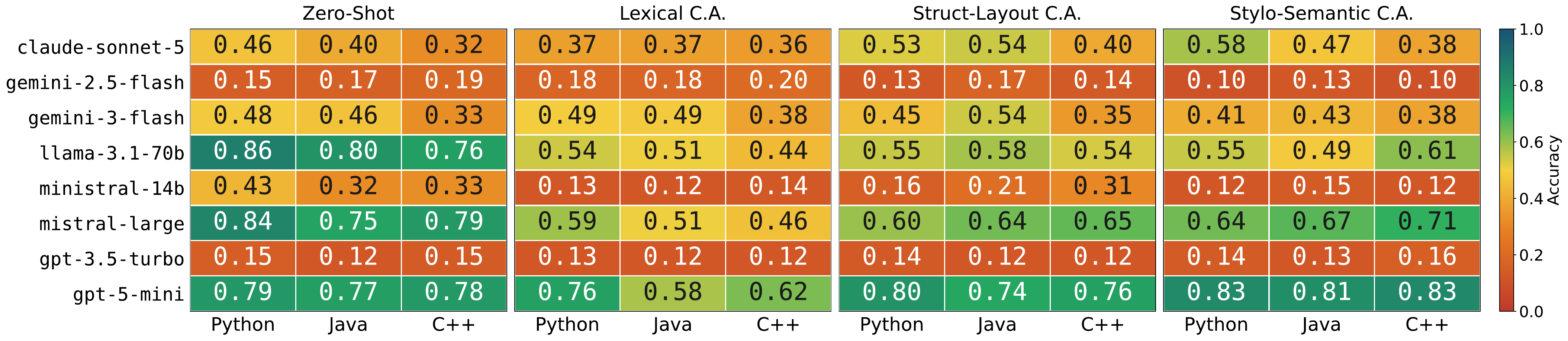}
         \caption{Accuracy}
         \label{fig:per_language_acc}
     \end{subfigure}
     \vfill
     \begin{subfigure}[b]{0.48\textwidth}
         \centering
         \includegraphics[width=\textwidth]{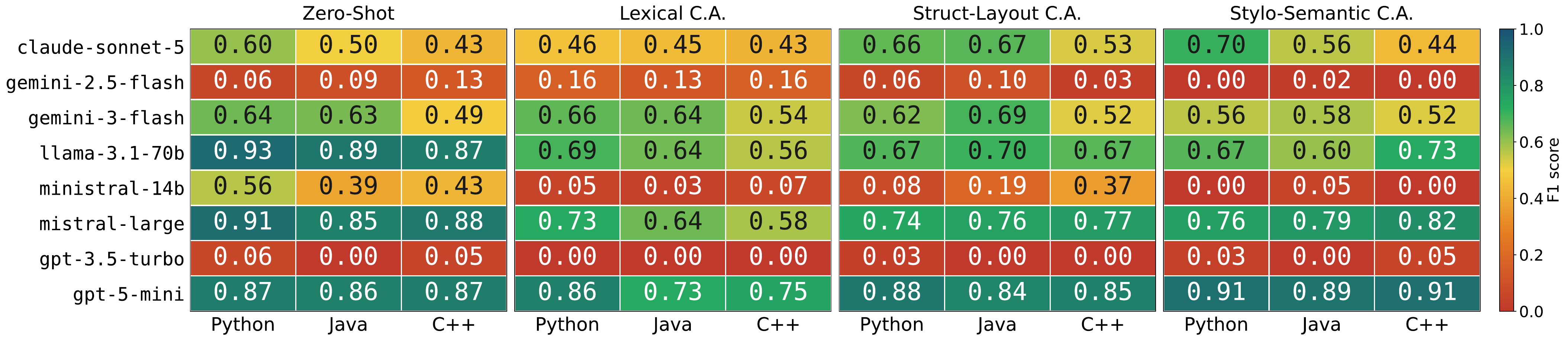}
         \caption{$F1_{score}$}
         \label{fig:per_language_f1}
     \end{subfigure}
    \caption{Per-language accuracy and $F1_{score}$ for each judge under the four prompting conditions.}
    \label{fig:per_language}
\end{figure}

Table~\ref{tab:unified-experience} and Figure \ref{fig:per_language} reports Accuracy/$F1_{score}$ (human-written as positive class) for all eight judges across the four prompting conditions and the three baselines, by generator. 

\begin{table}[t]
    \centering
    \caption{Accuracy/$F1_{score}$ by prompting condition and generator (human = positive class); bold = best per column.}
    \label{tab:unified-experience}
    \tiny
    \begin{tabular}{lp{0.6cm}p{0.6cm}p{0.6cm}p{0.6cm}p{0.6cm}p{0.6cm}p{0.6cm}p{0.6cm}}
        \toprule
        \textbf{Model / Approach} & \textbf{Star- Coder} & \textbf{Chat-GPT} & \textbf{BingAI Chat} & \textbf{Code Llama} & \textbf{Code Whispered} & \textbf{Github Copilot} & \textbf{Instruct Code T5+} \\
        \midrule
        \multicolumn{8}{c}{\textbf{Zero-shot}} \\
        \midrule
        Claude Sonnet 5 & 0.37/0.45 & 0.71/0.64 & 0.59/0.56 & 0.39/0.45 & 0.33/0.43 & 0.41/0.47 & 0.37/0.45 \\
        Gemini-2.5 & 0.53/0.68 & 0.56/0.69 & 0.50/0.67 & 0.50/0.67 & 0.50/0.67 & 0.53/0.68 & 0.53/0.68 \\
        Gemini-3 & 0.14/0.06 & 0.50/0.11 & 0.47/0.10 & 0.22/0.07 & 0.08/0.06 & 0.26/0.07 & 0.11/0.06 \\
        Llama & 0.50/0.00 & 0.50/0.00 & 0.47/0.00 & 0.47/0.00 & 0.44/0.00 & 0.42/0.00 & 0.44/0.00 \\
        Ministral & 0.50/0.55 & 0.50/0.55 & 0.36/0.49 & 0.50/0.55 & 0.50/0.55 & 0.47/0.54 & 0.47/0.54 \\
        Mistral Large & 0.56/0.20 & 0.53/0.19 & 0.44/0.17 & 0.50/0.18 & 0.53/0.19 & 0.39/0.15 & 0.50/0.18 \\
        GPT-3.5 & 0.50/0.67 & 0.50/0.67 & 0.50/0.67 & 0.50/0.67 & 0.53/0.68 & 0.50/0.67 & 0.56/0.69 \\
        GPT-5-mini & \textbf{0.82}/\textbf{0.82} & \textbf{0.96}/\textbf{0.95} & 0.77/0.78 & \textbf{0.82}/\textbf{0.82} & \textbf{0.92}/\textbf{0.90} & \textbf{0.76}/\textbf{0.75} & \textbf{0.85}/\textbf{0.82} \\
        \midrule
        \multicolumn{8}{c}{\textbf{Lexical Context Aware}} \\
        \midrule
        Claude Sonnet 5 & 0.56/0.68 & 0.94/0.94 & 0.69/0.74 & 0.51/0.65 & 0.49/0.64 & 0.52/0.67 & 0.53/0.67 \\
        Gemini-2.5 & 0.44/0.62 & 0.58/0.68 & 0.50/0.64 & 0.44/0.62 & 0.44/0.62 & 0.47/0.63 & 0.47/0.63 \\
        Gemini-3 & 0.17/0.12 & 0.56/0.20 & 0.50/0.18 & 0.31/0.14 & 0.08/0.11 & 0.31/0.14 & 0.28/0.13 \\
        Llama & 0.47/0.49 & 0.56/0.53 & 0.47/0.49 & 0.53/0.51 & 0.47/0.49 & 0.50/0.50 & 0.56/0.53 \\
        Ministral & 0.48/0.64 & 0.45/0.62 & 0.50/0.67 & 0.50/0.65 & 0.47/0.64 & 0.47/0.64 & 0.52/0.67 \\
        Mistral Large & 0.58/0.55 & 0.61/0.56 & 0.56/0.53 & 0.56/0.53 & 0.44/0.47 & 0.42/0.46 & 0.56/0.53 \\
        GPT-3.5 & 0.50/0.67 & 0.50/0.67 & 0.50/0.67 & 0.50/0.67 & 0.50/0.67 & 0.50/0.67 & 0.50/0.67 \\
        GPT-5-mini & 0.56/0.50 & 0.77/0.73 & 0.60/0.57 & 0.65/0.57 & 0.73/0.67 & 0.43/0.50 & 0.72/0.62 \\
        \midrule
        \multicolumn{8}{c}{\textbf{Structural-layout Context Aware}} \\
        \midrule
        Claude Sonnet 5 & 0.43/0.50 & 0.77/0.71 & 0.71/0.67 & 0.51/0.54 & 0.37/0.48 & 0.47/0.53 & 0.41/0.50 \\
        Gemini-2.5 & 0.50/0.65 & 0.47/0.64 & 0.47/0.64 & 0.47/0.64 & 0.47/0.64 & 0.47/0.64 & 0.47/0.64 \\
        Gemini-3 & 0.17/0.12 & 0.56/0.20 & 0.47/0.17 & 0.28/0.13 & 0.14/0.11 & 0.28/0.13 & 0.25/0.13 \\
        Llama & 0.69/0.70 & 0.44/0.57 & 0.56/0.62 & 0.72/0.72 & 0.67/0.68 & 0.61/0.65 & 0.81/0.79 \\
        Ministral & 0.56/0.68 & 0.47/0.64 & 0.47/0.64 & 0.58/0.69 & 0.56/0.68 & 0.56/0.68 & 0.56/0.68 \\
        Mistral Large & 0.64/0.55 & 0.64/0.55 & 0.53/0.48 & 0.56/0.50 & 0.44/0.44 & 0.44/0.44 & 0.58/0.52 \\
        GPT-3.5 & 0.50/0.67 & 0.50/0.67 & 0.50/0.67 & 0.50/0.67 & 0.53/0.68 & 0.50/0.67 & 0.50/0.67 \\
        GPT-5-mini & 0.61/0.46 & 0.86/0.67 & 0.67/0.50 & 0.67/0.46 & 0.79/0.60 & 0.59/0.46 & 0.70/0.46 \\
        \midrule
        \multicolumn{8}{c}{\textbf{Lexico-semantic Context Aware}} \\
        \midrule
        Claude Sonnet 5 & 0.55/0.65 & 0.87/0.88 & \textbf{0.79}/\textbf{0.82} & 0.53/0.65 & 0.53/0.67 & 0.55/0.67 & 0.61/0.68 \\
        Gemini-2.5 & 0.44/0.62 & 0.44/0.62 & 0.44/0.62 & 0.44/0.62 & 0.44/0.62 & 0.44/0.62 & 0.44/0.62 \\
        Gemini-3 & 0.26/0.32 & 0.67/0.50 & 0.47/0.39 & 0.39/0.38 & 0.17/0.29 & 0.37/0.35 & 0.25/0.31 \\
        Llama & 0.64/0.68 & 0.53/0.62 & 0.72/0.74 & 0.64/0.68 & 0.67/0.70 & 0.61/0.67 & 0.72/0.74 \\
        Ministral & 0.50/0.67 & 0.50/0.67 & 0.50/0.67 & 0.50/0.67 & 0.50/0.67 & 0.50/0.67 & 0.53/0.68 \\
        Mistral Large & 0.64/0.55 & 0.72/0.62 & 0.53/0.48 & 0.61/0.53 & 0.47/0.46 & 0.50/0.47 & 0.61/0.53 \\
        GPT-3.5 & 0.50/0.67 & 0.50/0.67 & 0.51/0.68 & 0.57/0.71 & 0.55/0.71 & 0.53/0.69 & 0.50/0.67 \\
        GPT-5-mini & 0.57/0.25 & 0.83/0.40 & 0.58/0.29 & 0.80/0.40 & 0.73/0.33 & 0.62/0.29 & 0.75/0.33 \\
        \midrule
        \multicolumn{8}{c}{\textbf{Existing Detectors}} \\
        \midrule
        DetectCodeGPT \cite{xu_detecting_2024} & 0.50/0.00 & 0.50/0.00 & 0.50/0.00 & 0.50/0.00 & 0.50/0.00 & 0.50/0.00 & 0.50/0.00 \\
        GPT-Sniffer \cite{nguyen_gptsniffer_2024} & 0.50/0.00 & 0.50/0.00 & 0.50/0.00 & 0.50/0.00 & 0.50/0.00 & 0.50/0.00 & 0.50/0.00 \\
        GPT-Zero & 0.55/0.27 & 0.58/0.28 & 0.58/0.28 & 0.52/0.26 & 0.58/0.28 & 0.55/0.27 & 0.55/0.27 \\
        \bottomrule
    \end{tabular}
\end{table}

\noindent
\textbf{Finding 1: Existing detectors exhibit a short half-life.}
DetectCodeGPT~\cite{xu_detecting_2024} and GPT-Sniffer~\cite{nguyen_gptsniffer_2024}, despite strong results in their original settings, collapse to Accuracy$=0.50$/$F1_{score}{=}0.00$ on every generator; GPT-Zero remains weak ($F1_{score}$ $0.26$--$0.28$). Their failure to transfer to this generator set demonstrates that a detector's useful life is bounded by changes in the generation distribution.

\noindent
\textbf{Finding 2: LLM judges are a promising but conditional alternative.}
GPT-5-mini reaches zero-shot $F1_{score}$ of $0.75$--$0.95$, far above the dedicated detectors. Feature guidance recovers Llama (up to $0.79$) and lifts Claude to $0.94$, but degrades GPT-5-mini to $0.25$--$0.40$ under lexico-semantic prompting. Adaptability therefore does not remove the need for judge- and prompt-specific validation.

\noindent
\textbf{Finding 3: Accuracy can mask biased detectors.}
The dedicated detectors' 0.50 accuracy looks like chance performance, but $F1_{score}{=}0.00$ reveals near-constant ``AI'' predictions. The same illusion appears in LLM judges: Llama predicts almost only ``AI,'' whereas Gemini-2.5 and GPT-3.5 lean toward ``human.'' Claude also ranges from above $0.82$ on ChatGPT and BingAI Chat to below $0.50$ on StarCoder. Accuracy alone therefore neither diagnoses class bias nor establishes transfer.

\noindent
\textbf{Finding 4: Per-language sensitivity is judge-specific.}
C++ is Claude's weakest language under every prompting condition, while GPT-5-mini stays largely language-invariant throughout; Llama and Mistral-Large both flip C++ from their \emph{worst} language under zero-shot/lexical prompting to their \emph{best} under structural/stylo-semantic guidance, showing that richer context can reshape which language a judge finds easiest rather than uniformly improving all of them.

\vspace{-0.3cm}

\section{Conclusion}
\label{sec:conclusion}

We examined whether general-purpose LLMs can serve as training-free judges of code provenance (RQ1) and how feature-guided prompting affects their reliability (RQ2). Zero-shot judging is highly model-dependent: GPT-5-mini achieves strong discrimination, while several judges collapse toward a single class despite near-chance accuracy. Feature-guided prompting is not uniformly beneficial: it recovers collapsed judges such as Llama but degrades the strongest zero-shot judge, GPT-5-mini. Judgments also vary markedly by generator, suggesting sensitivity to generator-specific style rather than a universal AI/human signal. LLM-based provenance judges should therefore be validated per model and generator, not selected from aggregate scores alone. Future work should examine hybrid human-AI code and provenance-preserving transformations.

\newpage
\section*{Data-Availability Statement}
\label{sec:data}
The data used in this research are derived from publicly available sources. To ensure transparency and reproducibility, the source code, datasets, and usage instructions are publicly accessible on Zenodo~\cite{djire2026llm4gcd_artifact}. 
\bibliographystyle{ACM-Reference-Format}
\bibliography{main}

@misc{li_resilient_2024,
	title = {Resilient {Watermarking} for {LLM}-{Generated} {Codes}},
	url = {http://arxiv.org/abs/2402.07518},
	urldate = {2024-03-10},
	publisher = {arXiv},
	author = {Li, Boquan and Zhang, Mengdi and Zhang, Peixin and Sun, Jun and Wang, Xingmei},
	month = feb,
	year = {2024},
	note = {arXiv:2402.07518 [cs]},
}

@misc{zan_large_2023,
	title = {Large {Language} {Models} {Meet} {NL2Code}: {A} {Survey}},
	shorttitle = {Large {Language} {Models} {Meet} {NL2Code}},
	url = {http://arxiv.org/abs/2212.09420},
	urldate = {2024-08-29},
	publisher = {arXiv},
	author = {Zan, Daoguang and Chen, Bei and Zhang, Fengji and Lu, Dianjie and Wu, Bingchao and Guan, Bei and Wang, Yongji and Lou, Jian-Guang},
	month = may,
	year = {2023},
	note = {arXiv:2212.09420 [cs]},
}

@inproceedings{pan_assessing_2024,
	title = {Assessing {AI} {Detectors} in {Identifying} {AI}-{Generated} {Code}: {Implications} for {Education}},
	issn = {2832-7578},
	shorttitle = {Assessing {AI} {Detectors} in {Identifying} {AI}-{Generated} {Code}},
	url = {https://ieeexplore.ieee.org/document/10554754/},
	doi = {10.1145/3639474.3640068},
	urldate = {2024-10-14},
	booktitle = {2024 {IEEE}/{ACM} 46th {International} {Conference} on {Software} {Engineering}: {Software} {Engineering} {Education} and {Training} ({ICSE}-{SEET})},
	author = {Pan, Wei Hung and Chok, Ming Jie and Wong, Jonathan Leong Shan and Shin, Yung Xin and Poon, Yeong Shian and Yang, Zhou and Chong, Chun Yong and Lo, David and Lim, Mei Kuan},
	month = apr,
	year = {2024},
	pages = {1--11},
}

@misc{ye_uncovering_2024,
	title = {Uncovering {LLM}-{Generated} {Code}: {A} {Zero}-{Shot} {Synthetic} {Code} {Detector} via {Code} {Rewriting}},
	shorttitle = {Uncovering {LLM}-{Generated} {Code}},
	url = {http://arxiv.org/abs/2405.16133},
	doi = {10.48550/arXiv.2405.16133},
	urldate = {2024-12-02},
	publisher = {arXiv},
	author = {Ye, Tong and Du, Yangkai and Ma, Tengfei and Wu, Lingfei and Zhang, Xuhong and Ji, Shouling and Wang, Wenhai},
	month = may,
	year = {2024},
	note = {arXiv:2405.16133},
}

@inproceedings{jianwang_empirical_2024,
	address = {Sacramento CA USA},
	title = {An {Empirical} {Study} to {Evaluate} {AIGC} {Detectors} on {Code} {Content}},
	isbn = {979-8-4007-1248-7},
	url = {https://dl.acm.org/doi/10.1145/3691620.3695468},
	doi = {10.1145/3691620.3695468},
	language = {en},
	urldate = {2024-12-02},
	booktitle = {Proceedings of the 39th {IEEE}/{ACM} {International} {Conference} on {Automated} {Software} {Engineering}},
	publisher = {ACM},
	author = {{JianWang} and Liu, Shangqing and Xie, Xiaofei and Li, Yi},
	month = oct,
	year = {2024},
	pages = {844--856},
}

@misc{shi_between_2024,
	title = {Between {Lines} of {Code}: {Unraveling} the {Distinct} {Patterns} of {Machine} and {Human} {Programmers}},
	shorttitle = {Between {Lines} of {Code}},
	url = {http://arxiv.org/abs/2401.06461},
	doi = {10.48550/arXiv.2401.06461},
	urldate = {2024-12-02},
	publisher = {arXiv},
	author = {Shi, Yuling and Zhang, Hongyu and Wan, Chengcheng and Gu, Xiaodong},
	month = jul,
	year = {2024},
	note = {arXiv:2401.06461},
}

@inproceedings{xu_detecting_2024,
	title = {Detecting {AI}-{Generated} {Code} {Assignments} {Using} {Perplexity} of {Large} {Language} {Models}},
	volume = {38},
	url = {https://ojs.aaai.org/index.php/AAAI/article/view/30361},
	number = {21},
	urldate = {2024-12-02},
	booktitle = {Proceedings of the {AAAI} {Conference} on {Artificial} {Intelligence}},
	author = {Xu, Zhenyu and Sheng, Victor S.},
	year = {2024},
	pages = {23155--23162},
}

@article{hou_large_2024,
	title = {Large {Language} {Models} for {Software} {Engineering}: {A} {Systematic} {Literature} {Review}},
	issn = {1049-331X, 1557-7392},
	shorttitle = {Large {Language} {Models} for {Software} {Engineering}},
	url = {https://dl.acm.org/doi/10.1145/3695988},
	doi = {10.1145/3695988},
	language = {en},
	urldate = {2024-12-02},
	journal = {ACM Transactions on Software Engineering and Methodology},
	author = {Hou, Xinyi and Zhao, Yanjie and Liu, Yue and Yang, Zhou and Wang, Kailong and Li, Li and Luo, Xiapu and Lo, David and Grundy, John and Wang, Haoyu},
	month = sep,
	year = {2024},
	pages = {3695988},
}

@misc{yang_zero-shot_2023,
	title = {Zero-{Shot} {Detection} of {Machine}-{Generated} {Codes}},
	url = {http://arxiv.org/abs/2310.05103},
	doi = {10.48550/arXiv.2310.05103},
	urldate = {2024-12-11},
	publisher = {arXiv},
	author = {Yang, Xianjun and Zhang, Kexun and Chen, Haifeng and Petzold, Linda and Wang, William Yang and Cheng, Wei},
	month = oct,
	year = {2023},
	note = {arXiv:2310.05103 [cs]},
}

@article{nguyen_gptsniffer_2024,
	title = {{GPTSniffer}: {A} {CodeBERT}-based classifier to detect source code written by {ChatGPT}},
	volume = {214},
	shorttitle = {{GPTSniffer}},
	url = {https://www.sciencedirect.com/science/article/pii/S0164121224001043},
	urldate = {2024-12-13},
	journal = {Journal of Systems and Software},
	publisher = {Elsevier},
	author = {Nguyen, Phuong T. and Di Rocco, Juri and Di Sipio, Claudio and Rubei, Riccardo and Di Ruscio, Davide and Di Penta, Massimiliano},
	year = {2024},
	pages = {112059},
}

@article{idrisov_program_2024,
	title = {Program {Code} {Generation} with {Generative} {AIs}},
	volume = {17},
	copyright = {http://creativecommons.org/licenses/by/3.0/},
	issn = {1999-4893},
	url = {https://www.mdpi.com/1999-4893/17/2/62},
	doi = {10.3390/a17020062},
	language = {en},
	number = {2},
	urldate = {2024-12-19},
	journal = {Algorithms},
	publisher = {Multidisciplinary Digital Publishing Institute},
	author = {Idrisov, Baskhad and Schlippe, Tim},
	month = feb,
	year = {2024},
	note = {Number: 2},
	pages = {62},
}

@inproceedings{mitchell2023detectgpt,
  title={Detectgpt: Zero-shot machine-generated text detection using probability curvature},
  author={Mitchell, Eric and Lee, Yoonho and Khazatsky, Alexander and Manning, Christopher D and Finn, Chelsea},
  booktitle={International conference on machine learning},
  pages={24950--24962},
  year={2023},
  organization={PMLR}
}

@inproceedings{skurla2026mcdok, title={mcdok at SemEval-2026 Task 13: Finetuning LLMs for Detection of Machine-Generated Code}, url={http://dx.doi.org/10.18653/v1/2026.semeval-1.47}, DOI={10.18653/v1/2026.semeval-1.47}, booktitle={Proceedings of the 20th International Workshop on Semantic Evaluation (2026)}, publisher={Association for Computational Linguistics}, author={Skurla, Adam and Macko, Dominik and Simko, Jakub}, year={2026}, pages={322–327} }

@inproceedings{yao2026coderipple, title={CodeRipple: Wavelet-Based Detection of LLM-Generated Code}, url={http://dx.doi.org/10.18653/v1/2026.acl-long.1777}, DOI={10.18653/v1/2026.acl-long.1777}, booktitle={Proceedings of the 64th Annual Meeting of the Association for Computational Linguistics (Volume 1: Long Papers)}, publisher={Association for Computational Linguistics}, author={Yao, Xingyu and Mao, Zhendong and Wang, Quan}, year={2026}, pages={38351–38364} }

@inproceedings{moon2026dont, title={Don’t Judge Code by Its Cover: Exploring Biases in LLM Judges for Code Evaluation}, url={http://dx.doi.org/10.18653/v1/2026.findings-eacl.70}, DOI={10.18653/v1/2026.findings-eacl.70}, booktitle={Findings of the Association for Computational Linguistics: EACL 2026}, publisher={Association for Computational Linguistics}, author={Moon, Jiwon and Hwang, Yerin and Lee, Dongryeol and Kang, Taegwan and Kim, Yongil and Jung, Kyomin}, year={2026}, pages={1364–1389} }

@dataset{djire2026llm4gcd_artifact,
  author    = {Djire, Alberick Euraste},
  title     = {{LLM4GCD : LLMs for Generated Code Detection}},
  year      = {2026},
  publisher = {Zenodo},
  doi       = {10.5281/zenodo.22248135},
  url       = {https://doi.org/10.5281/zenodo.22248135},
}

\end{document}